\documentclass[aps,showpacs,showkeys,eqsecnum,nofootinbib,multicol]{revtex4}
\renewcommand{\theequation}{\arabic{equation}}
\def\beq{\begin{equation}}
\def\eeq{\end{equation}}
\def\bea{\begin{eqnarray}}
\def\eea{\end{eqnarray}}

\def\pa{\partial}

\begin{document}
\title{Warped products and FRW cosmology}
\author{Soon-Tae Hong}
\email{soonhong@ewha.ac.kr} 
\affiliation{Department of Science Education,
Ewha Womans University, Seoul 120-750 Korea}
\date{\today}
\begin{abstract}
We study the Friedmann-Robertson-Walker cosmological model to investigate non-smooth 
curvatures associated with multiple discontinuities involved in the
evolution of the universe, by exploiting Lorentzian warped product scheme.  
Introducing double discontinuities occurred at the radiation-matter and matter-lambda 
phase transitions, we also discuss non-smooth features of the spatially flat 
Friedmann-Robertson-Walker universe.
\end{abstract}
\maketitle


\section{Introduction}

A warped product spacetime was introduced by Bishop and O'Neill long ago~\cite{bo}.  
The warped product scheme was later applied to general 
relativity~\cite{be96} and semi-Riemannian geometry~\cite{be79}.  Recently, this warped products 
were extended to multiply warped products with non-smooth metric~\cite{choi00}, and 
the Banados-Teitelboim-Zanelli (BTZ) black hole~\cite{hong03grg} and 
Reissner-Nordstr\"om-anti-de Sitter (RN-AdS) black hole~\cite{hong03math}.  The concept 
of the warped products was used in higher dimensional theories.  
For instance, the warped products were exploited in the Randall-Sundrum model in 
five-dimensions~\cite{rs1} and in the Kaluza-Klein supergravity theory in 
seven-dimensions~\cite{kksug}.

On the other hand, the standard big bang cosmological model based on the 
Friedmann-Robertson-Walker (FRW) spacetimes has led to the inflationary 
cosmology~\cite{guth} and nowadays to the M-theory cosmology with bouncing universes~\cite{seiberg}.  
These spacetimes are foliated by a special set of spacelike hypersurfaces such that each 
hypersurface corresponds to an instant of time. 

In this paper, as a cosmological model we will exploit the FRW spacetimes, 
which can be treated as a warped product manifold possessing 
warping function (or scale factor) with time dependence, to investigate the 
non-smooth curvature originated from the multiple discontinuities involved 
in the evolution of the universe. In particular, we will analyze the spatially 
flat FRW universe by introducing double
discontinuities occurred at the radiation-matter and matter-lambda
phase transitions in the astrophysical phenomenology.

\section{Warped products with single discontinuity}
\setcounter{equation}{0}
\renewcommand{\theequation}{\arabic{section}.\arabic{equation}}

In differential geometry, a multiply warped product manifold 
$(M=B\times_{f_1}F_1\times...\times_{f_n}F_{n}, g)$ is defined to consist of the 
Riemannian base manifold $(B, g_B)$ and fibers $(F_i,g_i)$ ($i=1,...,n$) 
associated with the Lorentzian metric,
\beq
g=\pi_{B}^{*}g_{B}+\sum_{i=1}^{n}(f_{i}\circ\pi_{B})^{2}\pi_{i}^{*}g_{i}
\label{g}
\eeq
where $\pi_B$, $\pi_{i}$ are the natural projections of $B\times F_1\times...\times F_n$
onto $B$ and $F_{i}$, respectively, and $f_{i}$ are positive warping functions.  
For the specific case of $(B=R,g_B=-d\mu^{2})$, the Lorentzian metric is rewritten as
\beq
g=-d\mu^{2}+\sum_{i=1}^{n}f_{i}^{2}g_{i}.
\label{gnew}
\eeq
It is well-known that the Randall-Sundrum model is described by the metric 
of the form
\beq
g=-N^{2}(t,y)dt^{2}+A^{2}(t,y)d\vec{x}^{2}+dy^{2},
\eeq
which can be thus regarded as the Lorentzian metric for the (4+1) higher 
dimensional warped product manifold.

Next, let $M=B\times_{f_1} F_1\times...\times_{f_n} F_n$ 
be a multiply warped products with Riemannian curvature tensor $R$ and flow vector field 
$U=\pa_{t}$. If $X$, $Y\in V(B)$, $U_i$, $V_i\in V(F_i)$ ($n=1,2,...,n$), $d_i={\rm dim}~F_i$, 
$f_i\in C^0(S)$ at a single point $p\in B$, and 
$S=\{p\}\times _{f_1} F_1\times\cdots\times_{f_n} F_n$, then we can obtain the Ricci 
components as follows
\begin{eqnarray} 
{\rm Ric}(X,Y)&=&-\sum_{i=1}^{n}d_i
X^{1}Y^{1}\frac{f_i''(t)+\delta(t-p)
\left({f_i'}^{+}-{f_i'}^{-}\right)}{f_i},\nonumber\\
{\rm Ric}(X,U_i)&=&0,\nonumber\\
{\rm Ric}(U_i,V_i)&=&{}^{F_{i}}{\rm Ric}(U_i, V_i)+\langle
U_i, V_i\rangle
\frac{f_i''(t)+\delta(t-p)({f_i'}^{+}-{f_i'}^{-})}{f_i}
\nonumber\\
& &+\langle U_i, V_i\rangle\left[
(d_{i}-1)\frac{{f_{i}'}^{+}-{f_{i}'}^{-}}{f_{i}^{2}}
+\sum_{j\neq i} d_{j}\frac{\langle {f_i'}^{+}-{f_i'}^{-},\ {f_j'}^{+}
-{f_j'}^{-}\rangle}{f_i f_j}\right],\nonumber\\
{\rm Ric}(U_i, U_j)&=&0,~~{\rm for}~~i\neq j,
\end{eqnarray}
where $X=X^{1}\partial/\partial_{t}$ and $Y=Y^{1}\partial/\partial_{t}$ and 
$\delta(t-p)$ is the delta function.

\section{Warped products with multiple discontinuities}
\setcounter{equation}{0}
\renewcommand{\theequation}{\arabic{section}.\arabic{equation}}

Now, in order to study multiple discontinuities in the warped products, 
we consider the FRW metric of the form
\beq
g=-dt^{2}+f^{2}(t)\left(\frac{dr^{2}}{1-kr^{2}}+r^{2}(d\theta^{2}+\sin^{2}\theta
d\phi^{2})\right),
\label{frwmetric}
\eeq
where $f$ is a scale factor and $k$ is a parameter denoting the spatially flat ($k=0$), 
3-sphere ($k=1$) and hyperboloid ($k=-1$) universes.  Treating $f$ as a warping 
function associated with the Lorentzian metric (\ref{frwmetric}), the FRW spacetime can be 
regarded as the warped product manifold $M=B\times_{f} F$ where the base manifold $B$ is 
an open interval of $R$ with usual metric $g_{B}=-dt^2$, the fiber is a 3-dimensional 
Riemannian manifold $(F, g_F)$ and the warping function $f$ is any positive function on 
$B$. The Lorentzian metric (\ref{frwmetric}) for the Friedman-Robertson-Walker spacetime 
is then rewritten as $g=-dt^2+f^2(t)g_F$.  Here the warping function $f$ is a function of time 
alone and it measures how physical separations change with time.  The dynamics of the expanding 
universe only appears implicitly in the time dependence of the warping function $f$.

Given a warped product manifold $M=B\times_{f} F$, we assume $f$ be a positive smooth 
function on $B=(t_{0},\ t_{\infty})$ such as $f\in C^{\infty}$ for $t\not=t_{i}$ and 
$f\in C^{0}$ at $t=t_{i}$ $(i=1,2,...,n)$.  When $f\in C^{1}$ at points $t_{i}\in (t_{0},\
t_{\infty})$ and $S=\{t_{i}\}\times_{f}F$, we define $f\in
C^{0}(S)$ as a collection of functions $\{f^{(i)}\}$ with
$f^{(i)}$ piecewisely defined on the intervals $t_{i}\leq t\leq
t_{i+1}$ $(i=0,1,2,...,n)$ with $t_{n+1}=t_{\infty}$. Since $f\in
C^{0}(S)$, we have $f^{(i-1)}=f^{(i)}$, but ${f^{(i-1)}}'={f^{(i)}}'$. We will use 
the unit step function $\mu$ for discontinuity of ${f^{(i)}}'$ at $t=t_i$.

In order to evaluate the Ricci curvatures, we derive $f^{\prime}$ in terms of the
collection of functions $\{f^{(i)}\}$ with $f^{(i)}$ piecewisely
defined on the intervals $t_{i}\leq t\leq t_{i+1}$.  For a single discontinuity $n=1$ case, 
$f^{\prime}$ is trivially given by
\beq
f^{\prime}=f^{(1)\prime}\mu(t-t_{1})+f^{(0)\prime}\mu(t_{1}-t),
\eeq 
with $\mu(t-t_i)$ being the unit step function.  For double discontinuities 
$n=2$ case, $f^{\prime}$ is similarly given by
\begin{eqnarray}
f^{\prime}&=&\left(f^{(2)\prime}-\frac{1}{2}f^{(1)\prime}
+\frac{1}{2}f^{(0)\prime}\right)\mu(t-t_{2})+\left(
\frac{1}{2}f^{(1)\prime}-\frac{1}{2}f^{(0)\prime}\right)\mu(t-t_{1})\nonumber\\
&&+\left(\frac{1}{2}f^{(1)\prime}+\frac{1}{2}f^{(0)\prime}\right)
\mu(t_{2}-t) +\left(-\frac{1}{2}f^{(1)\prime}
+\frac{1}{2}f^{(0)\prime}\right)\mu(t_{1}-t). \label{f2pp}
\end{eqnarray}
By using iteration method, one can obtain for an arbitrary $n$ case
\begin{eqnarray}
f^{\prime}&=&\left(f^{(n)\prime}-f^{(n-1)\prime}
+\frac{1}{n}\sum_{k=0}^{n-1}f^{(k)\prime}\right)\mu(t-t_{n})
+\sum_{l=1}^{n-1}\left(-f^{(l-1)\prime}
+\frac{1}{n}\sum_{k=0}^{n-1}f^{(k)\prime}\right)\mu(t-t_{l})\nonumber\\
&&+\frac{1}{n}\sum_{k=0}^{n-1}f^{(k)\prime}\mu(t_{n}-t)
+\sum_{l=1}^{n-1}\left(-f^{(l)\prime}
+\frac{1}{n}\sum_{k=0}^{n-1}f^{(k)\prime}\right)\mu(t_{l}-t).\label{fp}
\end{eqnarray}
To proceed to calculate $f^{\prime\prime}$, we use the derivative of the unit step function $\mu(t-t_i)$.  
For all $t\not=t_i$ this is well-defined, $\mu'(t-t_{i})=0$. However, at $t=t_i$ 
there exists a jump discontinuity so that we cannot define 
classical derivative and thus we use the $\delta$-function, $\mu'(t-t_i)=\delta(t-t_i)$ 
to yield 
\begin{eqnarray}
f^{\prime\prime}&=&\left(f^{(n)\prime\prime}-f^{(n-1)\prime\prime}
+\frac{1}{n}\sum_{k=0}^{n-1}f^{(k)\prime\prime}\right)\mu(t-t_{n})
+\sum_{l=1}^{n-1}\left(-f^{(l-1)\prime\prime}
+\frac{1}{n}\sum_{k=0}^{n-1}f^{(k)\prime\prime}\right)\mu(t-t_{l})\nonumber\\
&&+\frac{1}{n}\sum_{k=0}^{n-1}f^{(k)\prime\prime}\mu(t_{n}-t)
+\sum_{l=1}^{n-1}\left(-f^{(l)\prime\prime}
+\frac{1}{n}\sum_{k=0}^{n-1}f^{(k)\prime\prime}\right)\mu(t_{l}-t)\nonumber\\
&&+\left(f^{(n)\prime}-f^{(n-1)\prime}\right)\delta(t-t_{n})
+\sum_{l=1}^{n-1}\left(f^{(l)\prime}-f^{(l-1)\prime}\right)\delta
(t-t_{l}). \label{fpp}
\end{eqnarray}
We can then obtain the Ricci components
\begin{eqnarray}
{\text{Ric}}(U,U)&=&-{\frac{3{f}''}{f}} \nonumber\\
{\text{Ric}}(U,X)&=&0\nonumber\\
{\text{Ric}}(X,Y)&=&\left({\frac{2(f^{\prime
2}+k)}{f^2}}+{\frac{{f}''}{f}}\right)\langle X,\ Y\rangle,~~~~~
{\text{if}}\ \ X, Y\perp U,
\end{eqnarray}
and the Einstein scalar curvature 
\beq
R=6\left(\frac{f^{\prime
2}}{f^{2}}+\frac{f^{\prime\prime}}{f}+\frac{k}{f^{2}}\right),
\eeq
where $f^{\prime}$ and $f^{\prime\prime}$ are given by (\ref{fp}) and (\ref{fpp}).

\section{FRW universe as warped products}
\setcounter{equation}{0}
\renewcommand{\theequation}{\arabic{section}.\arabic{equation}}

In the spatially flat FRW cosmology with $k=0$, the early universe
was radiation dominated, the adolescent universe was matter
dominated, and the present universe is now entering into
lambda-dominate phase in the absence of vacuum energy. If the
universe underwent inflation, there was a very early period when
the stress-energy was dominated by vacuum energy.  The Friedmann
equation may be integrated to give the age of the universe in
terms of present cosmological parameters.  We have the scale
factor $f$ as a function of time $t$ which scales as $f(t)\propto
t^{1/2}$ for a radiation-dominated (RD) universe, and scales as
$f(t)\propto t^{2/3}$ for a matter-dominated (MD) universe, and
scales as $f(t)\propto e^{Kt}$ for a lambda-dominated (LD)
universe.  Note that the transition from the radiation-dominated phase 
to the matter-dominated is not an abrupt one; neither is the later
transition from the matter-dominated phase to the exponentially
growing lambda-dominated phase.  With the above astrophysical 
phenomenology in mind, we consider the spatially flat FRW spacetime $(M, g)$ 
with Lorentzian metric $g=-dt^2+f^2(t)g_{F}$ in the form of warped products 
as in (\ref{frwmetric}) with $k=0$.  Here $f$ is a smooth function on 
$B=(t_{0},\ t_{\infty})$ except at $t\not=t_{i}$ $(i=1,2)$, that is $f\in C^{\infty}(S)$ 
(where $S=\{t_{i}\}\times_{f}F$) for $t\not=t_{i}$ and $f\in C^{0}(S)$ 
at $t=t_{i}\in B$ to yield
\begin{equation}
f=\left(\begin{array}{l}
f^{(0)}=c_{0}t^{1/2},~~~~\mbox{for~$t<t_{1}$}\\
f^{(1)}=c_{1}t^{2/3},~~~~\mbox{for~$t_{1}\leq t\leq t_{2}$}\\
f^{(2)}=c_{2}e^{Kt},~~~~\mbox{for~$t>t_{2}$}
\end{array}
\right) \label{f012}
\end{equation}
with the boundary conditions
\begin{equation}
c_{0}t_{1}^{1/2}=c_{1}t_{1}^{2/3},~~~c_{1}t_{2}^{2/3}=c_{2}e^{Kt_{2}}.
\label{bc}
\end{equation}
Experimental values for $t_{1}$ and $t_{2}$ are given by
$t_{1}=4.7\times 10^{4}$ yr and $t_{2}=9.8$ Gyr~\cite{ry}.
Moreover $c_{1}$ and $c_{2}$ are given in terms of $c_{0}$,
$t_{1}$ and $t_{2}$ as follows
$$c_{1}=c_{0}t_{1}^{-1/6},~~~c_{2}=c_{0}t_{1}^{-1/6}t_{2}^{2/3}e^{-Kt_{2}}.$$
Note that in the spatially flat FRW model, $f\in C^0(S)$ since if
we assume $f\in C^1(S)$ one could have the boundary conditions
$\frac{1}{2}c_{0}t_{1}^{-1/2}=\frac{2}{3}c_{1}t_{1}^{-1/3}$ and
$\frac{2}{3}c_{1}t_{2}^{-1/3} =Kc_{2}e^{Kt_{2}}$, which cannot
satisfy the above boundary conditions (\ref{bc}) simultaneously.

Substituting $f$ in (\ref{f012}) into (\ref{fp})
and (\ref{fpp}), one can readily obtain 
\begin{eqnarray}
f^{\prime}&=&\left(\frac{1}{4}c_{0}t^{-1/2}-\frac{1}{3}c_{1}t^{-1/3}
+Kc_{2}e^{Kt}\right)\mu(t-t_{2})
+\left(-\frac{1}{4}c_{0}t^{-1/2}+\frac{1}{3}c_{1}t^{-1/3}
\right)\mu(t-t_{1})\nonumber\\
&&+\left(\frac{1}{4}c_{0}t^{-1/2}+\frac{1}{3}c_{1}t^{-1/3}\right)\mu(t_{2}-t)
+\left(\frac{1}{4}c_{0}t^{-1/2}-\frac{1}{3}c_{1}t^{-1/3}\right)\mu(t_{1}-t)\label{fpfrw}\\
f^{\prime\prime}&=&\left(-\frac{1}{8}c_{0}t^{-3/2}+\frac{1}{9}c_{1}t^{-4/3}
+K^{2}c_{2}e^{Kt}\right)\mu(t-t_{2})
+\left(\frac{1}{8}c_{0}t^{-3/2}-\frac{1}{9}c_{1}t^{-4/3}
\right)\mu(t-t_{1})\nonumber\\
&&+\left(-\frac{1}{8}c_{0}t^{-3/2}-\frac{1}{9}c_{1}t^{-4/3}\right)\mu(t_{2}-t)
+\left(-\frac{1}{8}c_{0}t^{-3/2}+\frac{1}{9}c_{1}t^{-4/3}\right)\mu(t_{1}-t)\nonumber\\
&&+\left(-\frac{2}{3}c_{1}t^{-1/3}+Kc_{2}e^{Kt}\right)\delta (t-t_{2})
+\left(-\frac{1}{2}c_{0}t^{-1/2}+\frac{2}{3}c_{1}t^{-1/3}\right)\delta
(t-t_{1}), \label{fppfrw}
\end{eqnarray}
with $\mu(t-t_i)$ and $\delta(t-t_{i})$ being the unit step
function and the delta function, respectively.

For the spatially flat FRW spacetime $M=B\times_{f}F$ with Riemannian curvature $R$, 
flow vector field $U=\partial_t$ and warping function $f\in C^0(S)$, we can then obtain 
the Ricci curvature for vector fields $X$, $Y$, $Z\in V(F)$, 
\begin{eqnarray}
{\text{Ric}}(U,U)&=&-{\frac{3{f}''}{f}} \nonumber\\
{\text{Ric}}(U,X)&=&0\nonumber\\
{\text{Ric}}(X,Y)&=&\left({\frac{2f^{\prime
2}}{f^2}}+{\frac{{f}''}{f}}\right)\langle X,\
Y\rangle,~~~~~{\text{if}}\ \ X, Y\perp U,
\end{eqnarray}
and the Einstein scalar curvature 
\beq
R=6\left(\frac{f^{\prime
2}}{f^{2}}+\frac{f^{\prime\prime}}{f}\right),
\eeq
where $f$, $f^{\prime}$ and $f^{\prime\prime}$ are given by
(\ref{f012}), (\ref{fpfrw}) and (\ref{fppfrw}), respectively.

\section{Conclusions}

We have considered the FRW cosmological model in the warped
product scheme to investigate the non-smooth curvature originated from 
the multiple discontinuities associated with the evolution of the
universe. In particular we have analyzed the non-smooth features
of the spatially flat FRW universe phenomenologically by introducing double
discontinuities occurred at the radiation-matter and matter-lambda
phase transitions.

\begin{center}
{\bf Acknowledgments}
\end{center}

STH would like to acknowledge financial support in part
from the Korea Science and Engineering Foundation Grant
(R01-2000-00015).

\end{document}